\documentclass[twocolumn,twoside,superscriptaddress,flushbottom,showkeys,a4paper,nofootinbib]{revtex4-1}

\usepackage{amssymb,amsmath}
\usepackage{graphicx}
\usepackage{epstopdf}
\usepackage{chemarr}
\usepackage{bm}

\newcommand{\UPDATE}[1]{#1}

\providecommand{\FLinT}{Center for Fundamental Living Technology, University of Southern Denmark, Campusvej 55, 5230 Odense M, Denmark}
\providecommand{\ICREA}{ICREA-Complex Systems Lab, Universitat Pompeu Fabra (GRIB), Dr Aiguader 80, 08003 Barcelona, Spain}
\providecommand{\SFI}{Santa Fe Institute, 1399 Hyde Park Road, Santa Fe NM 87501, USA}

\begin{document}

\title{On the Growth Rate of Non-Enzymatic Molecular Replicators}
\author{Harold Fellermann}
\affiliation{\FLinT}
\affiliation{\ICREA}
\email{harold@ifk.sdu.dk}
\author{Steen Rasmussen}
\affiliation{\FLinT}
\affiliation{\SFI}

\begin{abstract}
It is well known that non-enzymatic template directed molecular replicators $X + nO \stackrel k \longrightarrow 2X$ exhibit parabolic growth $d[X]/dt \propto k[X]^{1/2}$. Here, we analyze the dependence of the effective replication rate constant $k$ on hybridization energies, temperature, strand length, and sequence composition.
First we derive analytical criteria for the replication rate $k$ based on simple thermodynamic arguments.
Second we present a Brownian dynamics model for oligonucleotides that allows us to simulate their diffusion and hybridization behavior. The simulation is used to generate and analyze the effect of strand length, temperature, and to some extent sequence composition, on the hybridization rates and the resulting optimal overall rate constant $k$. 
Combining the two approaches allows us to semi-analytically depict a fitness landscape for template directed replicators. The results indicate a clear replication advantage for longer strands at low temperatures.
\end{abstract}

\keywords{non-enzymatic molecular replication; growth rate; product inhibition; reaction kinetics; Brownian dynamics}

\maketitle{}



\section{Introduction}

Optimizing \UPDATE{the yield of non-enzymatically self-replicating biopolymers} is of great interest for many basic science and application areas.
Clearly, the early organisms could not emerge with a fully developed enzymatic gene replication machinery, so it is plausible that the first organisms had to rely on non-enzymatic replication~\cite{Gil:1986,Mon:2008,Cle:2009}.
Most bottom up protocell models also rely on non-enzymatic biopolymer replication~\cite{Ras:2004,Ras:2004b,Ras:2008b,Szo:2001,Man:2008,Han:2009},
which is also true for \UPDATE{a variety of} prospective molecular computing and manufacturing applications.\footnote{For example, see the European Commission sponsored projects MatchIT and ECCell.} Common for all of these research areas is the interest to obtain an optimal replication yield in the absence of modern enzymes.
Depending on the details the biopolymer can be deoxyribonucleic acid (DNA), ribonucleic acid (RNA), peptide nucleic acid (PNA), etc.
In the following we'll refer to them as XNA. 

\begin{figure}
	\centering
	\includegraphics[width=.85\columnwidth]{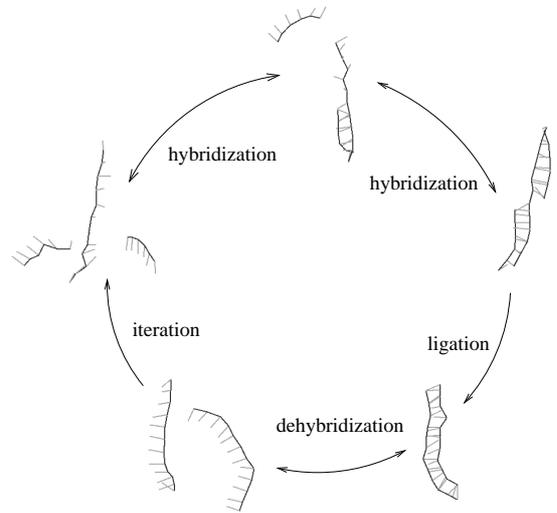}
	\caption{
		\UPDATE{Minimal template directed replicator: two complementary oligomers hybridize to a template strand (upper part). An irreversible ligation reaction transforms the oligomers into the complementary copy of the template. The newly obtained double strand can dehybridize (lower part) thus allowing for iteration of the process. We assume that ligation is rate limiting, which implies that hybridization and dehybridization are in local equilibrium.}
	}
	\label{fig_schematics}
\end{figure}

Conceptually, XNA replication proceeds in three basic steps:
(a) association, or \emph{hybridization} of \UPDATE{$n$} nucleotide monomers or oligomers with a single stranded, complementary template; (b) formation of covalent bonds in a condensation reaction, called \emph{polymerization} in case of monomer condensation and \emph{ligation} in case of oligomers; and finally (c) dissociation, or \emph{dehybridization} of the newly formed complementary strand:
\begin{equation}
	X + n \, O
	\underbrace{\xrightleftharpoons[k^-_\text O]{k^+_\text O}}_\text{(a)} X O_n
	\underbrace{\xrightleftharpoons[k^-_\text L]{k^+_\text L}}_\text{(b)} X\bar X + (n-1) \, W
	\underbrace{\xrightleftharpoons[k^+_\text T]{k^-_\text T}}_\text{(c)} X + \bar X + (n-1) \, W .
	\label{eq_repgen}
\end{equation}
Here, $X$ and $\bar X$ denote the template strand and its complement, $O$ denotes monomers/oligomers, $W$ is the leaving group of the condensation reaction, and
\begin{equation}
	K_i = \frac{k^+_i}{k^-_i} = e^{-\Delta G_i/k_\text BT} \quad i \in \{ \text O, \text L, \text T \}
	\label{eq_rates}
\end{equation}
are the equilibrium constants of the three reactions. Note that the left hand transition of reaction scheme~\eqref{eq_repgen} is an abbreviation of a multi-step process that accounts for all \UPDATE{$n$} individual oligomer hybridizations and dehybridizations, which is only partly captured by the net reaction process. 

The covalent condensation reaction is entirely activation limited. For nucleotide monophosphates, the leaving group corresponds to water which (due to its high concentration in aqueous solution) pushes the equilibrium to the hydrolyzed state. Product yields are significantly increased when using activated nucleotides, such as nucleotide triphosphate or imidazole.

\UPDATE{Due to its complex reaction mechanism, non-enzymatic XNA replication from monomers suffers from various complications, namely extension of sequences containing consecutive adenine and thymine nucleotides~\cite{Wu:1992,Wu:1992b}, as well as side reactions such as partial replication and random strand elongation~\cite{Fer:2007}. Consequently, experiments with activated nucleotides typically show little yield in aqueous solution, although results can be enhanced by employing surfaces (e.g. clay) or up-concentration in water-ice~\cite{Mon:2008, Mon:2010b}.}

Replication from short activated oligomers, on the other hand, does produce high yields for both RNA and DNA~\cite[and references therein]{Kie:1986,Sie:1994,Bag:1996,Joy:1984,Lin:2009}.
\UPDATE{In particular, this observation has lead to the development of \emph{minimal replicator systems}, in which ligation of two oligomers is sufficient to form the complementary replica \UPDATE{(see Fig.~\ref{fig_schematics})}.
One of the reasons why these systems outperform replicators that draw from monomers is that the above side reactions expectedly occur, if at all, only to a negligible extent.
}

Neglecting both the production of waste as well as the hydrolysis of the ligation product, but explicitly taking into account the individual oligomer associations, \UPDATE{minimal replicator systems} (here for the case of a self-complementary template) can be written as
\begin{equation}
	X + 2 O
	\xrightleftharpoons[k^-_\text O]{k^+_\text O} X O + O
	\xrightleftharpoons[k^-_\text O]{k^+_\text O} X O_2
	\xrightarrow{k_\text L} X_2
	\xrightleftharpoons[k^+_\text T]{k^-_\text T} 2 X .
	\label{eq_minrep}
\end{equation}

In this article, we first develop a theoretical expression for the template directed replication rate \UPDATE{of minimal replicator systems} as a function of strand length and temperature.
This analytical model provides transparent physical relations for how temperature, strand length- and composition impact the overall replication rate. 
We then present a 3D, implicit solvent, constrained Brownian Dynamics model for short nucleotide strands, i.e. strands with negligible secondary structures.
The model does not attempt to be (quantitatively) predictive. 
In particular, we do not attempt to calibrate interaction parameters to experimental data, which \UPDATE{prevents} any sequence prediction. On the contrary, it is our aim to demonstrate that much of the replication properties of oligonucleotides arises from rather general statistical physics. 
The simulation is used to measure diffusion coefficients, effective reaction radii, and hybridization rates and their dependence on temperature, strand length, and, to some extent, sequence information. 
\UPDATE{This allows} us to qualitatively obtain equilibrium constants $K_\text O$ \UPDATE{and} $K_\text T$ 
\UPDATE{and} to qualitatively sketch the effective replication rate~$k$ as a function of strand length and temperature.
\UPDATE{
Our analysis focuses on minimal replicator systems in the context of chemical replication experiments as employed in protocell research and manufacturing applications, where the researcher controls reactant concentrations as well as most experimental parameters. However, we also discuss the impact of our findings in the context of origin of life research, were possible side reactions cannot be neglected.
}

\section{Parabolic Growth and Replication Rate}
\label{sec_thermodynamics}
Following and extending the derivation of Refs.~\cite{Wil:1998,Roc:2006}, we assume that ligation is the rate limiting step. This translates into the following conditions for the rate constants:
\begin{equation}
\begin{array}{rlrlrl}
	k_\text L [XO_2] &\!\!\!\ll k_\text O^+ [X][O] \quad &
	k_\text L [XO_2] &\!\!\!\ll k_\text O^+ [XO][O] \\
	k_\text L [XO_2] &\!\!\!\ll k_\text O^- [XO] &
	k_\text L [XO_2] &\!\!\!\ll k_\text O^- [XO_2] \\
	k_\text L [XO_2] &\!\!\!\ll k_\text T^+ [X]^2 &
	k_\text L [XO_2] &\!\!\!\ll k_\text T^- [X_2]
\end{array}
\label{eq_rate_limits}
\end{equation}
One can then assume a steady state of the hybridization/dehybridization reactions and express the total template concentration
\(
	[X]_\text{total} = [X] + [XO] + [XO_2] + 2[X_2]
\)
in terms of equilibrium constants as
\begin{equation}
	[X]_\text{total} = \left(1 + K_\text O[O] + K_\text O^2[O]^2 \right)[X] + 2 K_\text T [X]^2 .
	\label{eq_X_total}
\end{equation}
When solved for $[X]$, this gives
\begin{multline}
	[X] = \frac{1}{4 K_\text T}\sqrt{8 K_\text T [X]_\text{total} + (1+K_\text O[O] + K_\text O^2[O]^2)^2} \\ - \frac{1+K_\text O[O] + K_\text O^2[O]^2}{4 K_\text T}.
	\label{eq_quadratic}
\end{multline}

Template directed replication typically suffers from product inhibition, where most templates are in double strand configuration, i.e. $K_\text T [X]_\text{total} \gg 1$. Over the course of the reaction, this is tantamount of saying that $\sqrt{8 K_\text T [X]_\text{total}} \gg 1+K_\text O[O] + K_\text O^2[O]^2$.
This allows us to approximate
\begin{multline}
	\sqrt{8 K_\text T [X]_\text{total} + (1+K_\text O[O] + K_\text O^2[O]^2)^2} \\ =
	\sqrt{8 K_\text T [X]_\text{total}} + \sqrt{(1+K_\text O[O] + K_\text O^2[O]^2)^2} \\
		+ \mathcal O([X]_\text{total})
\end{multline}
and simplify \eqref{eq_quadratic} to
\begin{equation}
 	    [X] = \sqrt{\frac{[X]_\text{total}}{{2 K_\text T}}} + \mathcal O([X]_\text{total}) .
	\label{eq_parabolic}
\end{equation}
This is a lower bound of the single strand concentration, which is approached in the limit of vanishing oligomer concentration.
By combining~\eqref{eq_minrep} and~\eqref{eq_parabolic}, we get
\begin{align}
	\frac{d[X]_\text{total}}{dt}
	&= k_\text L [XO_2]
	= k_\text L K_\text O^2 [O^2][X] \nonumber \\
	&\approx k \; [O]^2 \sqrt{[X]_\text{total}}
	\label{eq_total_growth}
\end{align}
\UPDATE{with}
\begin{equation}
k = k_\text L \frac{K_\text O^2}{\sqrt{2 K_\text T}} .
	\label{eq_total_rate}
\end{equation}
This well-established parabolic growth law is known to qualitatively alter evolutionary dynamics of XNA based minimal replicators \UPDATE{and to promote coexistence of replicators rather than selection of the fittest}~\cite{Sza:1989,Kie:1991}.\footnote{
\UPDATE{In particular, it has been shown that under parabolic growth conditions, competing replicators $X_i$ grow when sufficiently rare:
\[
	[X_i] < \left(\frac{k_i}{k_\text{base}} \frac{\sum_j [X_j]}{\sum_j [X_j]^{1/2}}\right)^2
	\Longrightarrow
	\frac{d}{dt} [X_i] > 0.
\]
The equation captures the connection between the growth rate $k_i$ and its selective pressure, such that replicator species with a high growth rate are also assigned a high evolutionary fitness. See Ref.~\cite{Sza:1989} for the derivation. It is this relation that allows us to speak of a \emph{fitness landscape} when referring to the functional shape of $k$.}
}
\UPDATE{Consequently, }several strategies have been designed to overcome product inhibition in order to reestablish Darwinian evolution \UPDATE{and survival of \emph{only} the fittest} \cite{Lut:1998,Roc:2006,Zha:2006,Lin:2009}. \UPDATE{Most of these approaches hinge on a mechanism to lower the hybridization tendency of the product to the template.} \UPDATE{In this article, however}, we accept parabolic growth and instead focus on the effective growth rate.

The key observation of equation~\eqref{eq_total_rate} is that, due to the steady state assumption, the overall growth rate is independent of the individual association and dissociation rates $k_i^+, k_i^-$, but only depends on the equilibrium constants $K_\text O$ and $K_\text T$. Expressed in free energy changes, Eq.~\eqref{eq_total_rate} becomes
\begin{equation}
	k = e^{\left[\log A + \left(\frac 1 2 \Delta G_\text T - 2 \Delta G_\text O - \Delta G^\ddag_\text L\right)\right]/k_\text BT},
	\label{eq_energies}
\end{equation}
where $A$ and $\Delta G^\ddag_\text L$ are the pre-exponential factor and activation energy of the ligation reaction, respectively, and we have used the Arrhenius equation
\begin{equation}
	k_\text L = A e^{-\Delta G^\ddag_\text L/k_\text BT} .
\label{eq_arrhenius}
\end{equation}

We further observe that any potential optimum of \eqref{eq_total_rate} must obey
\UPDATE{
\begin{align}
	2 {k_\text L}' K_\text T K_\text O &= \frac 1 2 k_\text L K_\text O {K_\text T}' - 4 k_\text L K_\text T K_\text O' \label{eq_max_rate}
\end{align}
}
where the prime indicates derivation with respect to any variable.
Note that derivatives of $k_\text L, K_\text T$, and $K_\text O$ can be taken with respect to parameters such as temperature and template length. In sequence space, however,
we do not have an ordering that would allow us to perform derivatives. Therefore, equation \eqref{eq_max_rate} can only give us partial information about an optimal growth rate.

It is well-known that the equilibrium constants $K_\text O$ and $K_\text T$ depend on various parameters such as temperature, salt concentration, strand length, and sequence information -- all being relevant control parameters when designing replication experiments or delimiting origin of life conditions~\cite{Owc:1998,Blo:2000}. 
Furthermore, the two rates are interdependent as one expects $K_\text T$ to \UPDATE{rise} with increasing $K_\text O$.

Qualitatively, the free energy of XNA hybridization obeys a form given by 
\begin{align}
	\Delta G(N,T) &= \UPDATE{N\Delta G_\text{base} + \Delta G_\text{init} \nonumber \\
	&= N (\Delta H_\text{base} - T \Delta S_\text{base}) \nonumber \\
	&\quad + \Delta H_\text{init} - T\Delta S_\text{init}} ,
	\label{eq_poland}
\end{align}
where \UPDATE{$N$ signifies the strand length, $\Delta G_\text{base}$ is the (maximal) energy change per base,} \UPDATE{$\Delta G_\text{init}$ is the initiation energy and} $\Delta H_\text{base}, \Delta S_\text{base}$ are negative\UPDATE{, whereas $\Delta H_\text{init}, \Delta S_\text{init}$ are positive}.
The right hand side of the equation expresses a saturation in the free energy per base as a function of the strand length; the free energy gain for each base pairing asymptotically becomes constant for long strands~\cite{Pol:1966b}. Inserting \eqref{eq_poland} into \eqref{eq_energies}
and separating out the rate constant for the ligation reaction $k_\text L$, we obtain:
\begin{align}
	\frac{K_\text O^2}{\sqrt{2K_\text T}}
	&\propto e^{{\left(\dfrac 1 2 \Delta G(N,T) - 2 \Delta G(N/2,T) \right)/k_\text BT}} \nonumber \\
	&\propto e^{- \left( \dfrac 1 2  N\Delta G_\text{base} + \dfrac 3 2 \Delta G_\text{init} \right) / k_\text BT } , \label{eq_Keff}
\end{align}
which, when differentiated for $T$, yields a positive dependence on temperature, iff
\begin{align*}
	\frac{d}{d\mathrm T} \frac{K_\text O^2}{\sqrt{2K_\text T}} > 0
	\quad\Longleftrightarrow\quad
	& \UPDATE{N < - \frac{3\Delta H_\text{init}}{\Delta H_\text{base}}}
\end{align*}

Since \UPDATE{$\Delta H_\text{init} \!>\! 0$, and $\Delta H_\text{base} \!\le\! 0$}, this critical strand length is truly positive.
It might surprise that $K_\text O^2/\sqrt{2K_\text T}$ can increase with decreasing temperature -- the regime where templates are primarily inhibited by the product. The results become understandable when considering that oligomers, with their lower hybridization rate, barely associate with the template if the temperature is raised.

Reintroducing the ligation reaction, this relation gets refined to
\begin{align}
k &= k_\text L \frac {K_\text O^2}{\sqrt{2K_\text T}} \nonumber \\
  &= e^{\log A - \left( \dfrac 1 2 N \Delta G_\text{base} + \dfrac 3 2 \Delta G_\text{init}  + \Delta G^\ddag_\text L \right) / k_\text BT } ,
\label{eq_k_analytic}
\end{align}
with the critical strand length
\begin{equation}
	\frac{dk}{d\mathrm T} > 0
	\quad\Longleftrightarrow\quad
	N < N^* = \UPDATE{- \frac{3 \Delta H_\text{init} + 2\Delta H^\ddag_\text L}{\Delta H_\text{base}}} .
	\label{eq_N_critical}
\end{equation}
In words: we can identify a critical strand length $N^*$ above which the overall replication rate $k$ increases with decreasing temperature. This critical strand length is determined by the hybridization enthalpies $\Delta H_\text{base}, \Delta H_\text{init}$, and activation enthalpy change $\Delta H^\ddag_\text L$ of ligation..

\begin{figure}
	\centering
	\includegraphics[width=\columnwidth]{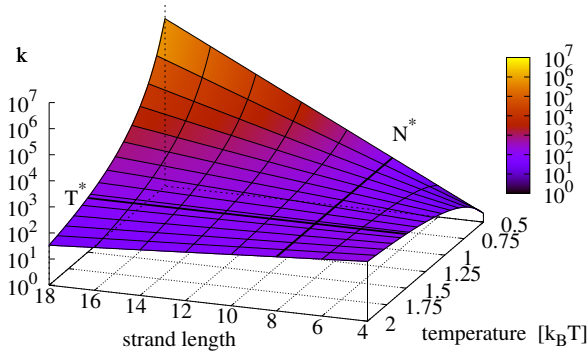}
	\caption{
		Effective replication rate $k$ (given by equation~\ref{eq_k_analytic}) as a function of strand length and temperature.
		For strands below a critical length $N^*$ (here 10) the rate increases with temperature, for strands longer than $N^*$, the replication rate grows with decreasing temperature. The value of $N^*$ is determined through equation~\eqref{eq_N_critical}.
		Note the saddle point of the surface where $T^*$ and $N^*$ intersect (Eq.~\ref{eq_max_rate})
		\UPDATE{($\Delta H_\text{base} \!=\! -1.5 k_\text BT'$, $\Delta S_\text{base} \!=\! -1 k_\text B$, $\Delta H_\text{init} \!=\! 0.50 k_\text BT'$, $\Delta S_\text{init} \!=\! 1.25 k_\text B$, $\Delta H^\ddag_\text{L} \!=\! 5.25 k_\text BT'$, $A = 10^3$)
}
	}
	\label{fig_analytic}
\end{figure}

Fig.~\ref{fig_analytic} depicts the graph of the replication rate landscape~\eqref{eq_k_analytic} that clearly identifies the optimum of equation~\eqref{eq_max_rate} as a saddle point.
The corresponding temperature $T^*$ where $k$ changes its scaling with respect to strand length is -- independent of the ligation reaction -- given by
\begin{equation}
	\frac{dk}{d\mathrm N} > 0
	\quad\Longleftrightarrow\quad
	T < T^* = \frac{\Delta H_\text{base}}{\Delta S_\text{base}} .
\label{eq_T_critical}
\end{equation}

Can we obtain a higher replication rate by using non-symmetric oligomers? The rational behind this strategy is to increase the binding affinity of one oligomer to maybe decrease product inhibition. A simple refinement of equation~\eqref{eq_Keff} allows us to capture this approach with our model:
\begin{align}
	\frac {K_{\text O_1} K_{\text O_2}}{\sqrt{2K_\text T}}
	&\propto e^{{\left(\frac 1 2 \Delta G(N,T) - \Delta G(N_1,T) - \Delta G(N_2,T) \right)/k_\text BT}} \nonumber \\
	&\propto e^ {\left(\UPDATE{\left(\frac 1 2 N - N_1 - N_2\right)}\Delta G_\text{base} - \frac  3 2 \Delta G_\text{init} \right)/k_\text BT}
\end{align}
where $N_1+N_2=N$ denote the lengths of oligomer strands $O_1$ and $O_2$. Thus, according to our simple thermodynamic considerations, non-symmetric variants of the replication process do not show more yield than the corresponding symmetric system: the binding affinity gained for the long oligomer strand is paid to hybridize the short oligomer strand.

\UPDATE{Fig.~\ref{fig_analytic} seemingly implies that replication rates grow beyond any limit for long templates, which is unphysical.
To resolve this inconsistency, it is important to remember that our findings are only valid in the regime where ligation is rate limiting.
For very long XNA strands, however, double strands are so stable that dehybridization of the ligation product is expected to become the rate limiting step.
Independent of the exact shape of the growth law, the dominant factor of the effective growth rate is given by
\begin{equation}
	k_\text T^- = k^+ \,e^{(N\Delta G_\text{base}+\Delta G_\text{init})/k_\text BT}
\end{equation}
where $k^+$ summarizes both pathways of either product rehybridization or hybridization of oligomers followed by ligation. As $k^+$ is composed of hybridization (i.e. diffusion plus orientational alignment) and ligation events, it varies only slightly with sequence length when compared to dehybridization rates for the case of large $N$. Therefore, the effective replication rate will be governed by the scaling
\begin{equation}
	k \propto e^{N\Delta G_\text{base}/k_\text BT}
	\label{eq_limit_N}
\end{equation}
with the limit
\[
	\lim_{N\rightarrow\infty}k=0,
\]
since $\Delta G_\text{base}<0$.
}
\UPDATE{As a consequence, we expect a full non-equilibrium study of the replication process to show a proper maximum in the replication rate as a function of strand length.}

\section{Spatially Resolved Replicator Model}
\label{sec_brownian}
Spatially resolved template-directed replicators have been previously simulated in the Artificial Life community using two-dimensional cellular automata and continuous virtual physics~\cite{Hut:2002,Smi:2003,Fer:2007}.  
The model we present here is conceptually similar to, but simpler than other coarse-grained DNA models~\cite[e.g.,][]{Kle:1998,Tep:2005,Dru:2000}.
Compared to our earlier work on hybridization and ligation~\cite{Fel:2007b}, the model presented here is less computationally expensive while simultaneously being broader in its range of application. 

We model nucleic acid strands as chains of hard spheres that are connected by rigid bonds. Each sphere has mass $m$, radius $r$, position and velocity $(\mathbf x_i,\mathbf v_i) \in \mathbb R^{3\times 3}$, as well as moment of inertia $\theta$, orientation and angular momentum $(\boldsymbol \omega_i, \mathbf L_i) \in \mathbb S^2 \times \mathbb R^3$ representing the spatial orientation of the respective nucleotide. Further, each sphere has a type $t_i \in \{\text A,\text T,\text C,\text G\}$, and we define A and T (C and G) to be complementary.
The model is implicit in the sense that solvent molecules are not represented explicitly, but only through their effect on the nucleotide strands.
We model the (translational and rotational) motion of each sphere by a \emph{Langevin equation}
\begin{subequations}
\begin{align}
	\stackrel . {\mathbf x}_i &= \mathbf v_i  \label{eq_langevin_pos} \\
	m \stackrel . {\mathbf v}_i &= - \boldsymbol \nabla U_i(\mathbf x, \boldsymbol \omega)
		- \gamma \mathbf v_i + \boldsymbol \xi_i \label{eq_langevin_v} \\
	\stackrel . {\boldsymbol \omega}_i &= \frac 1 \theta \; \mathbf L_i \times \boldsymbol \omega_i \\
	\stackrel {.} {\mathbf L}_i &= - \boldsymbol \nabla \hat U_i(\mathbf x, \boldsymbol \omega)
		- \gamma \frac{\mathbf L_i} \theta + \; \boldsymbol{\hat \xi}_i \label{eq_langevin_torque} .
\end{align}
\end{subequations}
Here, $\gamma$ is the friction coefficient, and $\boldsymbol \xi, \boldsymbol{\hat \xi}$ are zero mean random variables accounting for thermal fluctuations.
Together, friction and thermal noise act as a thermostat: they equilibrate the kinetic energy with an external heat bath whose temperature is given by the following \emph{fluctuation-dissipation-theorem}~\cite{Kub:1966}:
\begin{subequations}
\begin{align}
	\left<\boldsymbol \xi_i(t) ; \boldsymbol \xi_j(t')\right> &= 2\gamma k_\text B T \; \delta_{ij}\delta(t-t') \label{eq_fluc_diss_a} \\
	\left<\boldsymbol{\hat \xi}_i(t) ; \boldsymbol{\hat \xi}_j(t')\right> &= 2\frac\gamma\theta k_\text B T \; \delta_{ij}\delta(t-t') . \label{eq_fluc_diss_b}
\end{align}
\end{subequations}
Hence, a temperature change directly translates into a change of the Brownian noise amplitude.
We use the moment of inertia for solid spheres $\theta = \frac 2 5 m r^2$ -- noting that one could, in principle, use moment of inertia tensors to reflect the geometry of the individual nucleobases.

\begin{figure}
	\centering
	\includegraphics[width=0.9\columnwidth]{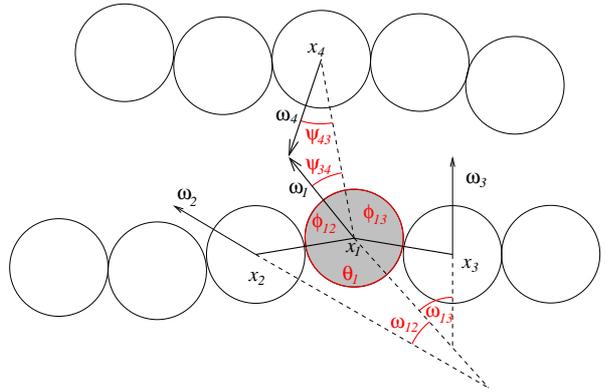}
	\caption{
		Geometry of the nucleotide strands. The figure shows the angles that define inner- and intermolecular
		interactions for one nucleobase (shaded in grey).
	}
	\label{fig_geometry}
\end{figure}

Equations~\eqref{eq_langevin_pos} - \eqref{eq_langevin_torque} are solved under the constraints
\begin{subequations}
\begin{align}
	\left| \mathbf x_i - \mathbf x_j \right| &= r_\text{bond} \quad \text{if $i, j$ bonded} \label{eq_constraint_bonds} \\
	\left| \mathbf x_i - \mathbf x_j \right| &= 2 r \quad \text{if } \left| \mathbf x_i - \mathbf x_j\right| < 2 r \: \label{eq_constraint_hard_sphere} \\
	\left| \boldsymbol \omega_i \right| &= 1 \label{eq_constraint_orientation}
\end{align}
\end{subequations}
to account for rigid bonds~\eqref{eq_constraint_bonds} and hard spheres~\eqref{eq_constraint_hard_sphere}.
By setting $r_\text{bond} < 2r$, we can assert that strands do not penetrate each other.
We define the following angles (see Fig.~\ref{fig_geometry}):
\begin{align*}
	\cos \theta_i = \left<\frac{\mathbf x_j-\mathbf x_i}{r_\text{bond}} \cdot \frac{\mathbf x_k-\mathbf x_i}{r_\text{bond}} \right> &\quad \text{$i,j$ and $i,k$ bonded} \\
	\cos \phi_{ij} = \left<\frac{\mathbf x_j-\mathbf x_i}{r_\text{bond}} \cdot \boldsymbol \omega_i\right> &\quad \text{$i,j$ bonded} \\
	\cos \omega_{ij} = \left<\boldsymbol \omega_i\cdot\boldsymbol \omega_j\right> &\quad \text{$i,j$ bonded} \\
	\cos \psi_{ij} = \left<\frac{\mathbf x_j-\mathbf x_i}{\left|\mathbf x_i-\mathbf x_j\right|}\cdot\boldsymbol\omega_i\right> &\quad \text{$i,j$ not bonded.}
\end{align*}

As much of the molecular geometry is already determined through the constraints, the innermolecular potentials $U$ and $\hat U$ only need to account for strand stiffness~\eqref{eq_U_bend}, \UPDATE{base orientation} \eqref{eq_U_ortho}, \UPDATE{and $\pi$-stacking} \eqref{eq_U_parallel}. We set:
\begin{subequations}
\begin{align}
	U^\text{bend}_i &= a_\text{bend} \UPDATE{\left(\dfrac{\theta_i}{\pi} - 1\right)^2} \label{eq_U_bend} \quad \text{if $i$ not terminal	} \\
	\hat U^\text{ortho}_{ij} &= \hat a_\text{ortho} \UPDATE{\left(\dfrac{\phi_{ij}}{\pi} - \dfrac 1 2\right)^2} \label{eq_U_ortho} \\
	\hat U^\text{parallel}_{ij} &= \hat a_\text{parallel} \UPDATE{\left(\dfrac{\omega_{ij}}{\pi} - 0\right)^2} \label{eq_U_parallel} .
\end{align}
The minimum energy state of these definitions are stretched out nucleotide strands with orientations perpendicular to the strand and parallel to each other.

In addition, we define the following intermolecular potentials between non-bonded complementary nucleobases $i$ and $j$:
\begin{align}
	U^\text{hybrid}_{ij} &= - a_\text{hybrid} \; d\left(\left|\mathbf x_i - \mathbf x_j\right|\right) \cos\psi_{ij} \label{eq_U_hybrid} \\
	\hat U^\text{hybrid}_{ij} &= -\hat a_\text{hybrid} \; d\left(\left|\mathbf x_i - \mathbf x_j\right|\right) \UPDATE{\left(\frac{\psi_{ij}}{\pi} - 1\right)^2} \label{eq_Uhat_hybrid} ,
\end{align}
if $\left|\mathbf x_i-\mathbf x_j\right| < r_\text c$.
The shift and weighing function
\[
	d(r_{ij}) = \frac 1 2 \left[ \cos\left(\frac{r_{ij}-2r}{r_\text c-2r}\pi\right) +1 \right]
\]
asserts that the potentials take on a minimum at particle contact and level out to zero at the force cutoff radius $r_\text c$.
Equation~\eqref{eq_U_hybrid} allows for a nucleobase $i$ to attract its complement $j$ along the direction of $\boldsymbol \omega_i$, while \eqref{eq_Uhat_hybrid} orients $\boldsymbol \omega_i$ toward the complement.

Taking the above potentials together, we define
\begin{align}
	U_i(\mathbf x, \boldsymbol \omega) &= U^\text{bend}_i(\mathbf x) + \!\!\!\!\!\!\!\! \sum_{\substack{i,j \\ \text{non-bonded} \\ \text{complementary} }} \!\!\!\!\!\!\!\! U^\text{hybrid}_{ij}(\mathbf x, \boldsymbol \omega) \\
	\hat U_i(\mathbf x, \boldsymbol \omega) &= \!\! \sum_{\substack{ i,j \\ \text{bonded} }} \!\! \left( \hat U^\text{ortho}_{ij}(\mathbf x, \boldsymbol \omega) + \hat U^\text{parallel}_{ij}(\boldsymbol \omega) \right) \nonumber \\
		&\quad + \!\!\!\!\!\!\!\! \sum_{\substack{i,j \\ \text{non-bonded} \\ \text{complementary} }} \!\!\!\!\!\!\!\! \hat U^\text{hybrid}_{ij}(\mathbf x, \boldsymbol \omega)
\end{align}
\end{subequations}

Equations~\eqref{eq_langevin_pos} to \eqref{eq_constraint_orientation} are numerically integrated using a Velocity Verlet algorithm that, in each iteration, first computes unconstrained coordinates which are afterwards corrected with a \textsc{Shake} algorithm to satisfy the constraints~\cite{Ryc:1977}.\footnote{Note that our approach would not work in the absence of a thermostat: to describe rotational motion properly, one would need to define orientations and angular momenta in a local reference frame that moves with the extended object to which the oriented point particle belongs. In this manner, rotational motion of the extended object gets propagated down to the angular momenta of the particles it consists of (A \textsc{QShake} algorithm would in addition be needed to properly conserve angular momenta in the constraints). While this approach is computationally significantly more cumbersome, we expect the result to be similar for the above model, in which rotation of extended objects is propagated down to its constituting particles through angular potentials and an overdamped thermostat.
}
\UPDATE{Typical system configurations are shown in Fig.~\ref{fig_schematics}.}

\begin{table}
\begin{center}
{\footnotesize
\begin{tabular}{|c|c|l|c|}
\hline
parameter		& value	& comment & Eqs. \\
\hline
$m$				& $1$	& particle mass & \eqref{eq_langevin_v} - \eqref{eq_langevin_torque} \\
$\gamma$		& $3$	& friction coefficient & \eqref{eq_langevin_v}, \eqref{eq_langevin_torque} \\
$k_\text BT_0$	& $1$	& equilibrium temperature & \eqref{eq_fluc_diss_a}, \eqref{eq_fluc_diss_b} \\
$\Delta t$		& $0.05$ & numerical time step & \\
\hline
$r$				& $0.25$	& particle radius & \eqref{eq_constraint_hard_sphere} \\
$r_\text{bond}$	& $0.45$	& bond length & \eqref{eq_constraint_bonds} \\
$r_\text c$		& $1$	& force cutoff radius & \eqref{eq_U_hybrid} - \eqref{eq_Uhat_hybrid} \\
\hline
$a_\text{bend}$				& $\UPDATE{5}$		& strand stiffness & \eqref{eq_U_bend} \\
$\hat a_\text{ortho}$		& $\UPDATE{2.5}$	& angular stretching & \eqref{eq_U_ortho} \\
$\hat a_\text{parallel}$	& $\UPDATE{1}$		& angular alignment & \eqref{eq_U_parallel} \\
$\hat a_\text{hybrid}$		& $\UPDATE{10}$		& angular hybridization & \eqref{eq_Uhat_hybrid} \\
$a_\text{hybrid}$			& $\UPDATE{1}$		& complementary attraction & \eqref{eq_U_hybrid} \\
\hline
\end{tabular}
}
\caption{
	Model parameters in reduced units (unless otherwise noted).
}
\label{table_parameters}
\end{center}
\end{table}

\section{Simulation Results}
\label{sec_results}

In the subsequent analyses, we will employ reduced units, i.e., $m=1$, $r_\text c=1$, and $k_\text BT_0=1$ define the units of mass, length, and energy. From this, the natural unit of time follows as
\[
	\tau = r \sqrt{m/k_\text BT_0}.
\]

\UPDATE{The parameters $r$ and $r_\text{bond}$ are chosen to prevent crossing of strands ($r_\text{bond} < 2r$). The ratio $r_\text{bond}/r$ determines the double strand geometry which is modeled more sparse than in actual nucleic acid strands in order to compensate for the relatively shallow potentials of the coarse-grained model. The ratio $r/r_\text c$ determines the distance at which complementary bases ``feel each other'' and has been set to two times the bead diameter.}

\UPDATE{The amplitudes $a_\text{bend}$, $\hat a_\text{ortho}$, $\hat a_\text{parallel}$ of the potential functions are chosen in order to promote stacked single strands for temperatures up  to at least $3 k_\text BT_0$ and to loosely match the persistence length of DNA (three nucleotides) at unit temperature.
Finally, the values for $\hat a_\text{hybrid}$ and $a_\text{hybrid}$ are chosen to enable hybridization at unit temperature.
We point out that our model utilizes a high value $\hat a_\text{hybrid}$ in order to promote fast hybridization.}
A list of all model parameters (unless otherwise noted) is given in Table~\ref{table_parameters}.

\UPDATE{Note that it is not mandatory to relate one bead of the model to one physical nucleotide. Instead, each bead could also represent a short XNA subsequence (e.g. 2-4 nucleotides). While this would result in a closer match of the ratio $r_\text{bond}/r$, the amplitudes of the potential functions would need to be adapted to reflect the changed representation.}

\subsection{Diffusion}
In dilute solution, DNA diffusion depends primarily on temperature and strand length, as opposed to primary or secondary structure.
In the limit of low Reynolds numbers, the diffusion coefficient of a sphere is given by the \emph{Einstein-Stokes equation}
\begin{equation}
	D = \frac{k_\text BT}{6\pi\eta r}
	\label{eq_einstein_stokes}
\end{equation}
where $\eta$ is the viscosity of the medium and $r$ the radius of the sphere.

\begin{figure}
	\centering
	\includegraphics[width=\columnwidth]{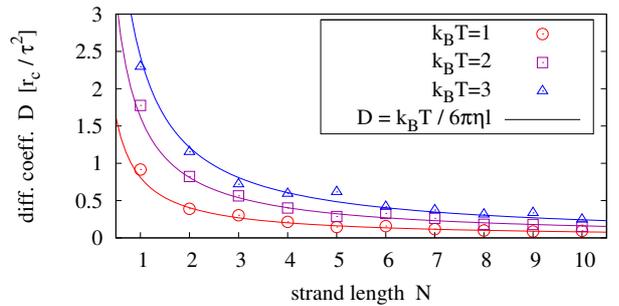}
	\caption{
		Diffusion coefficients measured for different strand lengths and temperatures (symbols) fitted to the prediction of the Einstein-Stokes relation (solid lines). For each parameter pair, 40 simulation runs over $1000\tau$ have been averaged.
	}
	\label{fig_diffusion}
\end{figure}

In order to compare our model polymer diffusion to \eqref{eq_einstein_stokes}, we perform simulations of single homopolymers (e.g. poly-C) and determine the diffusion coefficient from its measured mean square displacement
\[
	D=\frac 1 6 \frac{\left|\mathbf x(\Delta t)-\mathbf x(0)\right|^2}{\Delta t}
\]
Fig.~\ref{fig_diffusion} shows results for strands of lengths $N=1, \ldots, 10$ and temperatures $k_\text BT=1, \ldots, 3$.
The data sets a scaling relation between our model parameter $N$ and the Stokes radius $r$ which is \emph{a priori} not known.
For the general scaling relation $r\propto N^{\nu}$ we obtain the most likely exponent from data fitting via $\nu$ and $\eta$ as $1.06$.
By setting $\nu=1$, and equivalently $r\propto N$, we obtain the best agreement between measurement and theory (by fitting via $\eta$ only) for $\eta=0.061 k_\text BT\tau^2/r_\text c^2$ (see solid lines in Fig.~\ref{fig_diffusion}).

\subsection{Radius of gyration}

\begin{figure}
	\centering
	\includegraphics[width=\columnwidth]{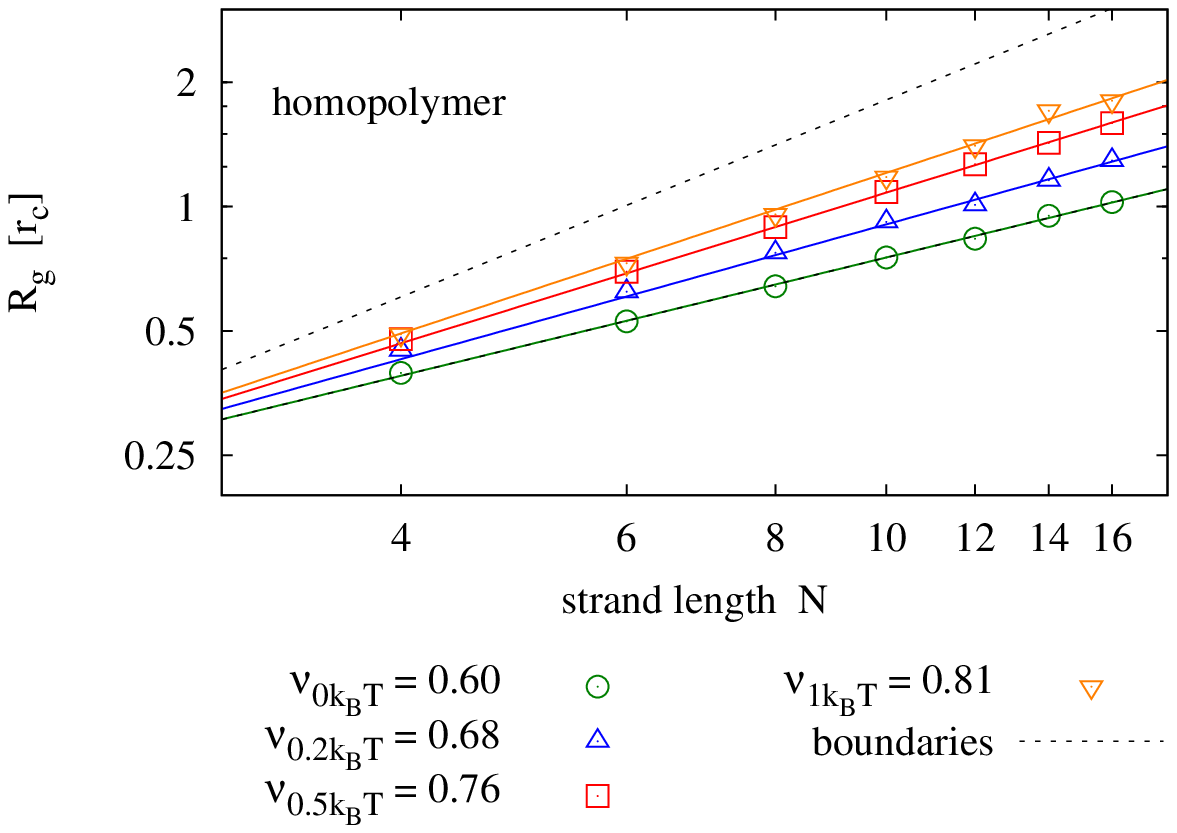}
	\includegraphics[width=\columnwidth]{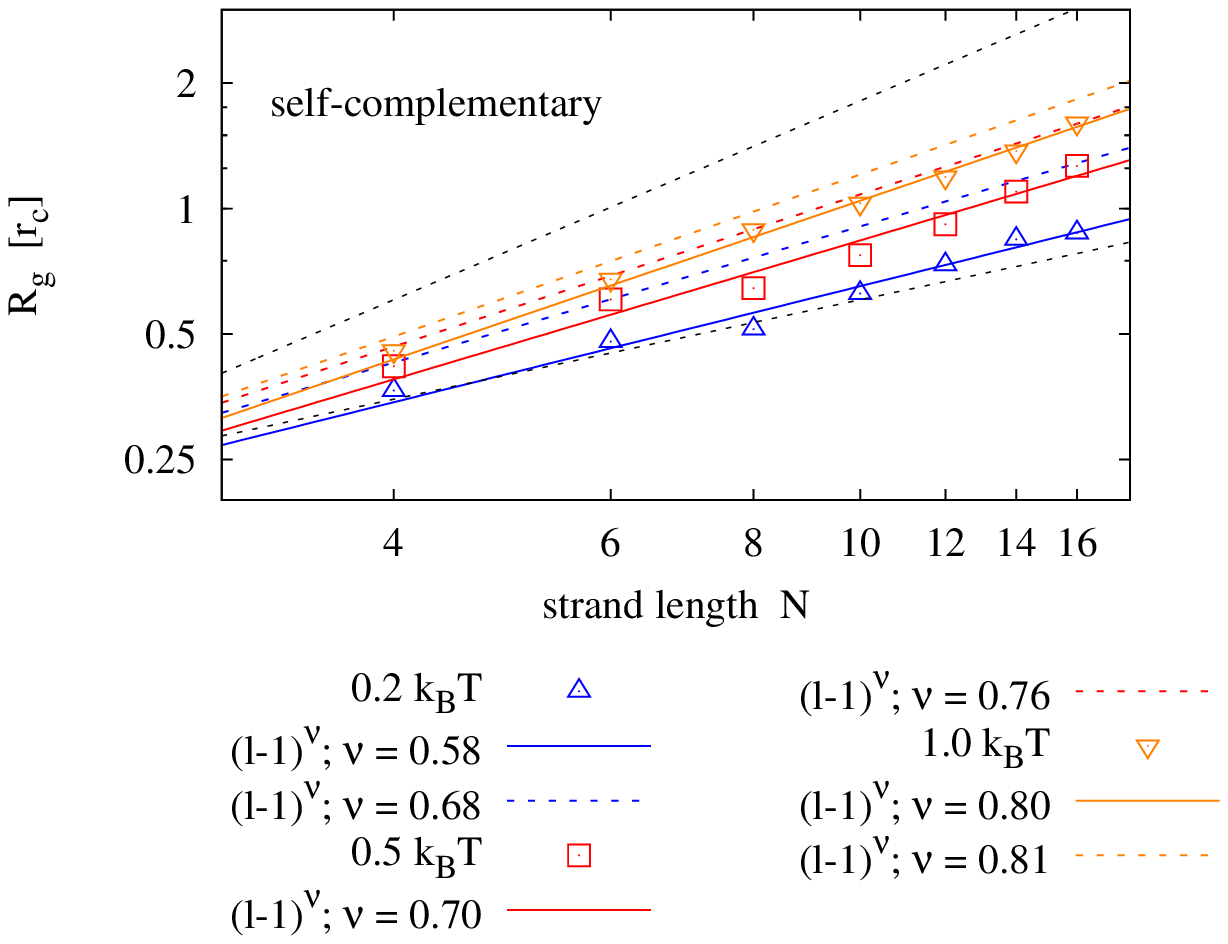}
	\caption{
		Radius of gyration measured for different strand lengths and bending potentials (symbols) fitted to the prediction of the Flory mean field theory (solid lines). For each parameter pair, 40 simulation runs over $400\tau$ have been averaged. The upper panel shows results for homopolymers (e.g. poly-C), the lower panel compares those to radii of self-complementary strands. The plots also show the boundaries for maximally stretched chains ($\nu=1$ -- upper dotted line) and the expectation value of an ideal chain ($\nu=\UPDATE{3/5}$ -- lower dotted line).
	}
	\label{fig_gyration}
\end{figure}

Again in dilute solution, the radius of gyration
\[
	R_\text g^2 = \frac 1 {N-1} \sum_{i=1}^{N}\left|\mathbf x_i - \mathbf x_\text{mean}\right|^2,
\]
with $\mathbf x_\text{mean}$ being the center of gravity of the chain, is expected to depend on chain length and temperature (or equally the backbone stiffness $a_\text{bend}$). As opposed to diffusion, we do expect the radius of gyration to change with the primary structure of the nucleotide strand.
For homopolymers, we expect $R_\text g$ to be well described by the Flory mean field model \cite{Ter:2002}
\[
	R_\text g \propto (N-1)^\nu .
\]

We perform simulations of single homopolymers and self-complementary nucleotide strands and determine the radius of gyration.
Fig.~\ref{fig_gyration} shows results for strands of lengths $N=4, \ldots, 16$ and various backbone stiffness values.
It is found that the Flory model is a good prediction, not only for homopolymers, but also for self-complementary strands. Expectedly, the radius of gyration is smaller for self-complementary strands. For $a_\text{bend}=0.2$, we find the radius of gyration of self-complementary strands to be slightly longer than the radius of gyration of a homopolymer with half the length -- implying that the strand is almost always in a hairpin configuration. For stronger backbone stiffness values, the effect is reduced.

\subsection{Melting behavior}
\label{sec_melting}

\begin{figure}
	\centering
	\includegraphics[width=\columnwidth]{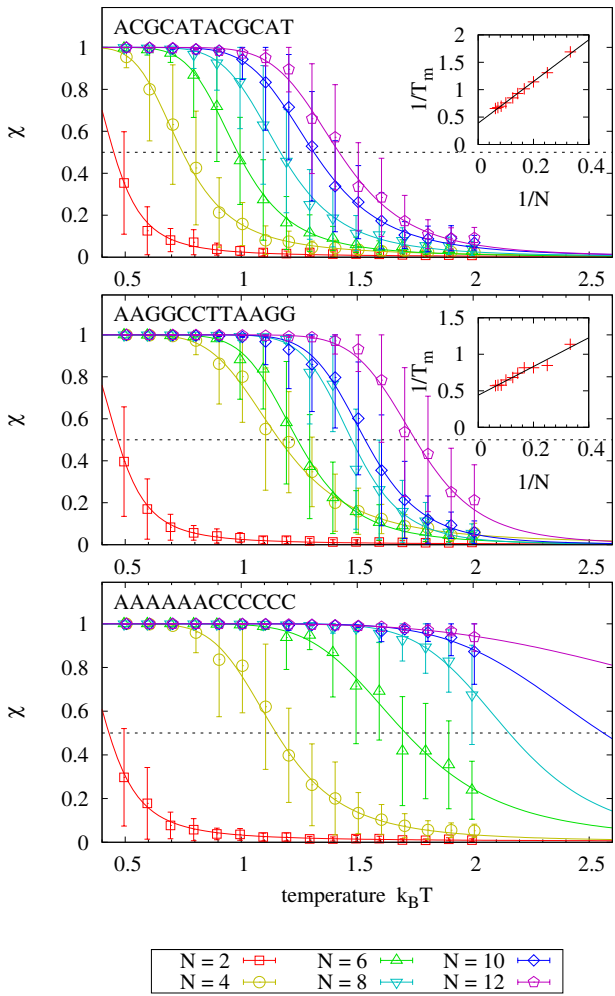}
	\caption{
		Systems of size $10^3$ are initialized with two complementary strands of length $N$. The sequence information is taken from the $N$ central nucleotides of \UPDATE{the} master sequence denoted in \UPDATE{each panel} (e.g., $N=6$ implies sequence CATACG in the first panel).  Each system is simulated over $\UPDATE{50000}\tau$, and \UPDATE{the average fraction $\chi$ of hybridized nucleobases} is determined. \UPDATE{Error bars show} the average and standard deviation of 40 measurements. Solid lines show the theoretical prediction $\chi(T) = \left(1 + e^{\frac{\Delta H - T\Delta S}{RT}} \right)^{-1}$ \UPDATE{fitted individually to each data set via $\Delta S$ and $\Delta H$}. \UPDATE{Melting temperatures $T_\text m$ are obtained from the relation $\chi(T_\text m)=0.5$, and their scaling as a function of strand length is depicted in the inlays for the cases where enough melting points had been observed.}
	}
	\label{fig_melting}
\end{figure}

We analyze the melting behaviour $[X]_2 \leftrightharpoons 2[X]$ of complementary nucleotide strands as a function of temperature for various strand lengths and sequences.
\UPDATE{We consider a base to be hybridized if there is a complementary base of another strand within a maximal distance of $r_\text c$.}
\UPDATE{Denoting the fraction of hybridized nucleobases with $0\le\chi\le 1$,
we can compare the melting curves to the theoretical prediction}
\begin{equation}
	\chi(T) = 
	\left(1 + e^{\frac{\Delta H - T\Delta S}{RT}} \right)^{-1}
	\label{eq_time_frac}
\end{equation}
where $\Delta H, \Delta S$ are constants depending on template length, sequence, and concentration.

Fig.~\ref{fig_melting}. shows melting curves for 18 different sequences and fits (via $\Delta H_i, \Delta S_i$) to the theoretical prediction, \UPDATE{where each panel analyzes sequences that are subsequences of a common master sequence denoted in each panel. The graphs clearly show how the average hybridization increases with strand length for each master sequence. Inlays, where present, emphasize that the inverse of the melting point, at which $\chi(T_\text m)=0.5$, scales linearly with the inverse of the strand length.}

\UPDATE{Comparing the individual panels to each other, we find that melting temperatures for strands of equal length are higher for sequences with identical adjacent bases.}
In fact, the melting behavior is dominated by the presence of identical adjacent bases: adding a single nucleobase to a strand that consists otherwise only of identical pairs (i.e., moving from length 4 to 6 and from length 8 to 10 in panel two) has no significant impact on the observed melting temperature.
We assume that this behavior is due to the fact that dehybridized nucleotides of a partly molten strand find more potential binding partners to enforce the stability of the partly molten strand, thereby promoting \UPDATE{recombination of products}. 

\UPDATE{We emphasize that our model is not suited to obtain quantitative sequence dependent melting temperatures. While it is well known that not only strand composition, but the actual arrangement of bases influences the melting temperature~\cite{San:1998}, the magnitude of this effect is not expected to be captured quantitatively by our model. Nevertheless, the results confirm that sequence information affects the melting temperature, and that the melting temperature rises when a strand is elongated.}

\begin{figure}
	\centering
	\includegraphics[width=\columnwidth]{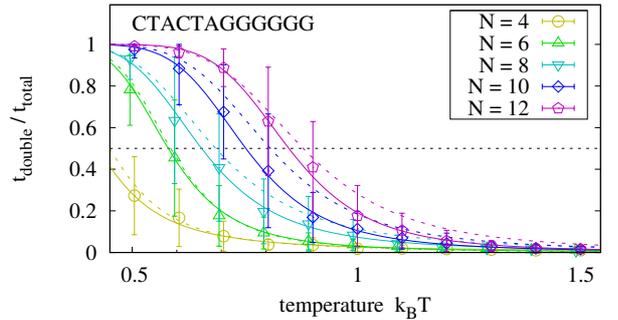}
	\caption{
		Melting curves for an oligomer that hybridizes to the left hand side of the master sequence in the presence of the right hand side oligomer. Data is obtained with the procedure described in Fig.~\ref{fig_melting}. For the analyzed master sequence, the results are comparable to those of two complementary strands of length $N/2$ (dotted lines).
	}
	\label{fig_oligomers}
\end{figure}

Up to now, we have analyzed hybridization of two complementary strands of equal length. How is the stability of the hybridization complex affected if one of the strands is replaced by two oligomers of half the length?
We analyze the master sequence CTACTAGGGGGG. Its left half is similar to the first sequence of Fig.~\ref{fig_melting} with respect to non-identical neighboring bases. The right half has been chosen for its strong hybridization tendency. We run experiments as before and measure the hybridization of the left oligomer. By comparing its equilibrium rate to the one for two templates of half the length, we can determine how the dangling right hand side affects the equilibrium rate (e.g., we compare the hybridization of a 4-mer to an 8-mer template to the hybridization of two 4-mers.)
Fig.~\ref{fig_oligomers} shows that the hybridization fractions $\chi_\text O(N)$ and $\chi_\text T(N/2)$ are comparable for the analyzed sequence.
We expect, however, that $\chi_\text O$ decreases when the two oligomers have more interaction possibilities than in the selected master sequence.

\subsection{Effective replication rate}
\label{sec_rep_rate}
We can roughly equate
\[
	\chi
	\equiv \frac{2[X_2]}{[X]_\text{total}}, \qquad
	1-\chi
	\equiv \frac{[X]}{[X]_\text{total}}
\]
and obtain an estimate for the equilibrium constant
\begin{equation}
	K_\text T = \frac{[X_2]}{[X]^2} \equiv \frac{\chi}{2(1-\chi)^2} \frac{1}{[X]_\text{total}} .
	\label{eq_hybr_rate}
\end{equation}
from the measurements.
This equation has to be taken with some caution because the measured hybridization times reflect a non-trivial relation between diffusing reactants and rehybridization of partly molten complexes -- both scaling differently with concentration. To truly obtain $K$, one is advised to repeat the simulations with varying concentrations, i.e. box size.
By means of \UPDATE{Eq.}~\eqref{eq_hybr_rate}, we convert the melting data from Sec.~\ref{sec_melting} to obtain hybridization energy changes 
\UPDATE{\begin{align*}
	\Delta G_\text T &= -k_\text BT\log K_\text T \\
	N \Delta G_\text{base} + \Delta H_\text{init} - T\Delta S_\text{init}  &= -k_\text BT \log\frac{\chi}{2(1-\chi)^2} \nonumber \\
	&\quad + k_\text BT\log [X]_\text{total}.
\end{align*}}
\UPDATE{In the latter equation, $-k_\text B \log{[X]_\text{total}}$ denotes the translational entropy for a box of size $[X]_\text{total}^{-1}$, while $\Delta S_\text{init}$ accounts for the configurational entropy of a single strand. We combine both entropy terms, $\Delta S' = \Delta S_\text{init} + k_\text B \log [X]_\text{total}^{-1}$, and fit
\begin{align*}
	N \Delta G_\text{base} + \Delta H_\text{init} - T\Delta S'  &= -k_\text BT \log\frac{\chi}{2(1-\chi)^2},
\end{align*}
}
\UPDATE{which allows for better comparison to the melting temperature plots, as
$\Delta G_\text T=0 \Longleftrightarrow \chi = 1/2$.}

\begin{figure}
	\centering
	\includegraphics[width=\columnwidth]{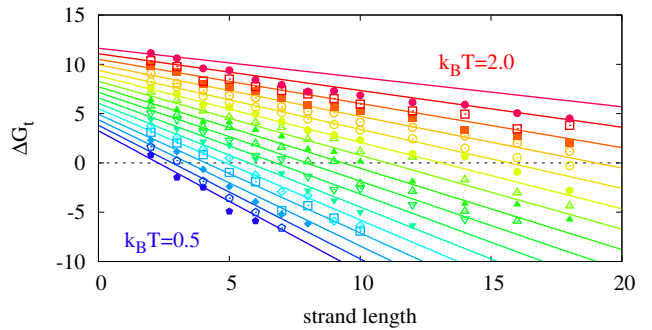}
	\caption{
		Hybridization energy changes $\Delta G_\text T$ obtained from the measurements of
		section~\ref{sec_melting}, sequence \UPDATE{ACGCATACGCATACGCATAC} (symbols), fitted 
		to the analytical model of equation
		\eqref{eq_poland} \UPDATE{via the parameters $\Delta H_\text{base} = -1.81$, $\Delta S_\textbf{•} = -0.756$, $\Delta H_\text{init} = 0.470$, and $\Delta S' = -5.58$. Since $[X]_\text{total} = 0.001$, we can estimate $\Delta S_\text{init}=1.33$.}
	}
	\label{fig_rates}
\end{figure}

\UPDATE{Determining hybridization energies is difficult because hybridization is very stable at low temperatures, particularly for long XNA strands. Dehybridization then becomes a rare event, which requires unfeasibly long simulation runs in order to sample equilibrium distributions. Consequently, data for low temperatures and long strands is overshadowed by noise and we have excluded such data from the analysis.}
\UPDATE{For the regime that is accessible to simulation, Fig.~\ref{fig_rates} shows the measured hybridization energies fitted to the theoretical model of Eq.~\eqref{eq_poland} -- see figure caption for details.
The data follows the linear trend of the model and can recover the proper temperature scaling.
However, we also observe deviations from the analytical prediction for $T>1.9$. 
The plot confirms that hybridization energy changes are close to zero at the melting temperature of each double strand. For the simulation, where $[X]_\text{total}=0.001$, we obtain $\Delta S_\text{init}=1.33$, which confirms that $\Delta G_\text{init}$ is primarily entropic.}

\UPDATE{Unfortunately, the removal of noisy simulation results implies that we do not have measurements for the regime where the analytical model predicts the most features. To nevertheless obtain estimates for these points, we perform the same simulations as before but start with a perfectly hybridized complex. For short strands, the difference in initial conditions is negligible, as hybridization and dehybridization equilibrate within the simulated time span. For long strands / strands at low temperatures, the sampled $\chi$ values progressively overestimate the equilibrium time fraction of hybridization.}

\UPDATE{The rational behind these dehybridization measurements is the following: for long strands and low temperatures
dehybridization of the ligation product becomes the rate limiting step and we are in the regime of Eq.~\eqref{eq_limit_N}. Here, the effective replication rate is primarily governed by the rate of product dehybridization which in turn gives us an upper bound for the replication rate.
By combining the results of the two scenarios, we implicitly relate the simulated time span with an assumed ligation rate.}

\begin{figure}
	\centering
	\includegraphics[width=\columnwidth]{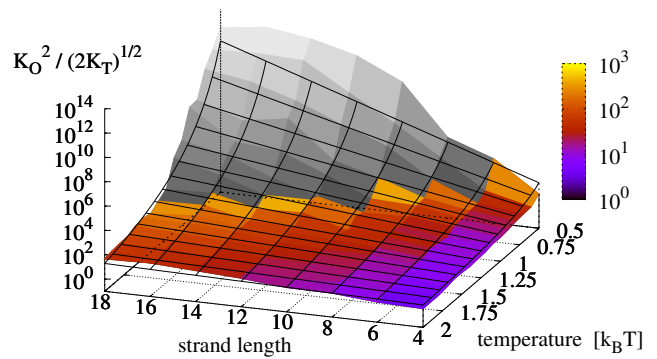}
	\caption{
		Effective equilibrium constant $K_\text O^2/\sqrt{2K_\text T}$, obtained
		from the measurements of \UPDATE{Fig.~\ref{fig_rates}} (colored) compared to the
		theoretical prediction of Eq.~\eqref{eq_poland} (mesh).
		\UPDATE{Data shaded in gray is extrapolated from dehybridization experiments.}
	}
	\label{fig_effective_rate}
\end{figure}

Plugging the measured constants $K_\text T$ and $K_\text O$ into \UPDATE{Eq.}~\eqref{eq_total_rate}, we obtain a fitness landscape for \UPDATE{minimal} replicators which is depicted in Fig.~\ref{fig_effective_rate}. 
\UPDATE{The colored surface shows the effective equilibrium constant $K_\text O^2/\sqrt{2K_\text T}$, obtained from the original measurements. The grey surface shows the results from the dehybridization experiments. Finally, the fitness landscape obtained via Eq.~\eqref{eq_poland} is shown as a mash.}
For the analyzed master sequence and range of observation, the effective oligomer complex concentration $K_\text O^2/\sqrt{2K_\text T}$ varies over \UPDATE{13} orders of magnitude with highest rates for long strands ($N\ge 8$) and low temperatures ($k_\text BT\le 0.8$).
\UPDATE{Contrary to the analytical derivations, the numerical results of the dehybridization experiments indicate a saturation and possibly a decrease of the replication rate for long strands at low temperatures, thereby supporting our hypothesis that the effective rate indeed possesses an optimum when dehybridization and ligation occurr on comparable time scales.}

\begin{figure}
	\centering
	\includegraphics[width=\columnwidth]{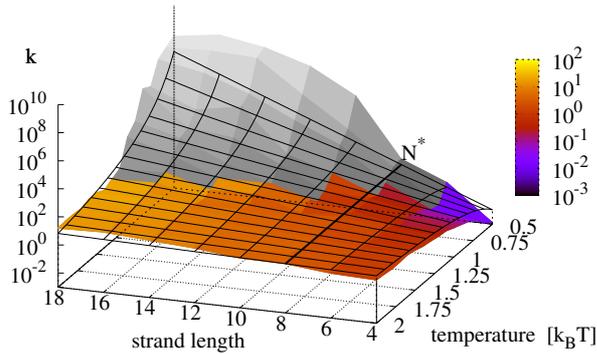}
	\caption{
		\UPDATE{Final replication rate $k$ as a function of template length and temperature. The figure is produced by superposing the data from Fig.~\ref{fig_effective_rate} with the Arrhenius equation for the ligation reaction following Eq.~\eqref{eq_k_analytic} with $A=10^3 ,\Delta H^\ddag_\text L=6.52 k_\text BT'$. For this parametrization, the critical strand length $N^*$ above which the temperature dependence of the reaction inverts is $8$.}
	}
	\label{fig_final_fitness}
\end{figure}

The numerical simulations do not incorporate the ligation reaction. To include its temperature dependence, we superpose the Arrhenius equation \eqref{eq_arrhenius} \UPDATE{with parameters as in Fig.~\ref{fig_analytic}} onto Fig.~\ref{fig_effective_rate} and obtain the replication rate landscape shown in Fig.~\ref{fig_final_fitness}. The resulting figure features a critical strand length \UPDATE{$N^* \approx 8$} at which the temperature scaling inverts. With the parameters obtained from the data fits, the critical temperature \UPDATE{$T^* \approx 2.37\;T_0$ lies outside the analyzed area}.

\section{Discussion}
The common strategy to increase the yield in template directed replication experiments is to increase the concentration of oligomers. This is certainly viable, and the fact that the growth rate $k$ is proportional to the square of the oligomer concentration encourages this approach. Our investigation, however, indicates that oligomer concentration can be outweighed drastically by factors such as temperature, template length, as well as sequence information, which all influence the replication rate at least exponentially and thus over many orders of magnitude. These finding are consistent for the simple analytical expressions (Fig.~\ref{fig_analytic}), for the simulations (Fig.~\ref{fig_rates}) as well as for the combined analysis (Figs.~\ref{fig_effective_rate} and \ref{fig_final_fitness}). 

Perhaps contrary to intuition, we find the highest growth rates for long replicators and low temperatures. This finding can be explained by the fact that the effective growth rate of minimal replicators features a critical strand length $N^*$ at which the temperature dependence of the overall replication reaction inverts: below $N^*$ the replication rate is dominated by the ligation reaction and its positive temperature scaling, whereas above $N^*$, the negative temperature scaling of the hybridization reactions becomes dominant, recall Fig.~\ref{fig_final_fitness}.

We observe that hybridization rates are highly sequence dependent. In particular, our spatially resolved simulations reveal that adjacent identical nucleobases can drastically stabilize the hybridization complex. We expect that the overall \UPDATE{replication process} is primarily sequence specific near to the ligation sites, as it is known that mismatches near the ligation site effect the ligation the most~\cite{Blo:2000}.

We also find that there is no difference in the replication rates of symmetric versus asymmetric replicators. We see that from equation 18, where only the sum of the oligomer lengths appear:\UPDATE{while the longer oligomer of an asymmetric replicator has a high binding affinity to the template and therefore promotes the formation of a hybridization complex, the short oligomer has a smaller binding affinity, such that the total asymmetric hybridization complex is as stable as its symmetric counterpart.}

We emphasize that our approach hinges on the assumption that ligation is the rate limiting step of the replication reaction. Due to the temperature scaling of the diffusion, hybridization, and ligation processes, our approach breaks down for very low temperatures or very long template strands. As discussed in Section \ref{fig_analytic}, equation~\eqref{eq_limit_N}, for long strands and low temperatures the dehybridization of the templates becomes the rate limiting step. Exactly what happens in the transition region between these two limits requires a more detailed non-equilibrium analysis and is outside the scope of this investigation.  The grey shaded area in Figs.~\ref{fig_effective_rate} and \ref{fig_final_fitness} depicts the expected landscape for the replication rate as we approach this transition zone from the regime where ligation is rate limiting, and it is clearly seen how the replication rate levels off as temperature decreases and the sequence length increases.  In any event, we would expect the existence of a true optimal temperature for a given strand length, and equally a true optimal strand length for a given temperature, such that replication rates are maximized.

In the context of origin of life research, where the temperature is given by the environment, but nucleic acid strands are subject to mutations, our findings suggest the existence of a critical temperature $T^*$, below which evolution would \UPDATE{select longer} nucleic acid replicators \UPDATE{that could have resulted from mutational elongations}, thereby promoting an increase of the potential information storage associated with these molecules. \UPDATE{Thus, these} results shed new light on the ``Snowball Earth'' hypothesis. \UPDATE{This argument, however, assumes that both templates and oligomers elongate to maintain the structure of a minimal replicator that replicates by ligation of two oligomers only. Needless to say, long oligomers of a specific sequence are less frequent in a random pool of substrate, such that they do not necessarily benefit from their increased stability.}

Most importantly, in the context of minimal replicator experiments and applications, e.g. in protocell as well as molecular computing \UPDATE{and fabrication} research, our findings suggest a qualitative recipe for obtaining high replication yields, as they relate \UPDATE{experimentally accessible data such as melting temperatures and ligation rate to the critical strand length (Eq.~\ref{eq_N_critical}) and temperature (Eq.~\ref{eq_T_critical})}.

\section*{Acknowledgements}
This work has benefited from discussions with the members of the FLinT Center for Fundamental Living Technology, University of Southern Denmark. In particular, we acknowledge P.-A. Monnard and C. Svaneborg, \UPDATE{as well as the anonymous reviewers of the journal \emph{Entropy}} for helpful feedback.


\bibliographystyle{mdpi}
\makeatletter
\renewcommand\@biblabel[1]{#1. }
\makeatother
\bibliography{references}

\begin{thebibliography}{10}

\bibitem{Gil:1986}
Gilbert, W.
\newblock The RNA world.
\newblock {\em Nature} {\bf 1986}, {\em 319},~618.

\bibitem{Mon:2008}
Monnard, P.A.
\newblock The dawn of the RNA world: RNA Polymerization from
  monoribonucleotides under prebiotically plausible conditions. In {\em
  Prebiotic Evolution and Astrobiology}; Wong, J.T.F.; Lazcano, A., Eds.;
  Landes Bioscience: Austin, TX, USA,  2008.

\bibitem{Cle:2009}
Cleves, J.H.
\newblock Prebiotic chemistry, the premordial replicator and modern protocells.
  In {\em Protocells: Bridging nonliving and living matter}; Rasmussen, S.;
  Bedau, M.; Chen, L.; Deamer, D.; Krakauer, D.; Packard, N.; Stadler, P.,
  Eds.; MIT Press,  2009; p. 583.

\bibitem{Ras:2004}
Rasmussen, S.; Chen, L.; Deamer, D.; Krakauer, D.C.; Packard, N.H.; Stadler,
  P.F.; Bedau, M.A.
\newblock Transitions from nonliving to living matter.
\newblock {\em Science} {\bf 2004}, {\em 303},~963--965.

\bibitem{Ras:2004b}
Rasmussen, S.; Chen, L.; Stadler, B.M.R.; Stadler, P.F.
\newblock Proto-Organism Kinetics: Evolutionary Dynamics of Lipid Aggregates
  with Genes and Metabolism.
\newblock {\em Orig. Life Evol. Biosph.} {\bf 2004}, {\em 34},~171--180.

\bibitem{Ras:2008b}
Rasmussen, S.; Bailey, J.; Boncella, J.; Chen, L.; Collis, G.; Colgate, S.;
  DeClue, M.; Fellermann, H.; Goranovic, G.; Jiang, Y.; Knutson, C.; Monnard,
  P.A.; Mouffouk, F.; Nielson, M.; Sen, A.; Shreve, A.; Tamulis, A.; Travis,
  B.; Weronski, P.; Zhang, J.; Zhou, X.; Ziock, H.J.; Woodruff, W.
\newblock Assembly of a minimal protocell. In {\em Protocells: Bridging
  Nonliving and Living Matter}; Rasmussen, S.; Bedau, M.; Chen, L.; Deamer, D.;
  Krakauer, D.; Packard, N.; Stadler, P., Eds.; MIT Press: Cambridge, USA,
  2008; pp. 125--156.

\bibitem{Szo:2001}
Szostack, W.; Bartel, D.P.; Luisi, P.L.
\newblock Synthesizing life.
\newblock {\em Nature} {\bf 2001}, {\em 409},~387--390.

\bibitem{Man:2008}
Mansy, S.S.; Schrum, J.P.; Krishnamurthy, M.; Tob{\'{e}}, S.; Treco, D.A.;
  Szostak, J.W.
\newblock Template-directed synthesis of a genetic polymer in a model
  protocell.
\newblock {\em Nature} {\bf 2008}, {\em 454},~122--125.

\bibitem{Han:2009}
Hanczyc, M.
\newblock Steps towards creating a synthetic protocell. In {\em Protocells:
  Bridging Nonliving and Living Matter}; Rasmussen, S.; Bedau, M.; Chen, L.;
  Deamer, D.; Krakauer, D.; Packard, N.; Stadler, P., Eds.; MIT Press:
  Cambridge, USA,  2009; p. 107.

\bibitem{Wu:1992}
Wu, T.; Orgel, L.E.
\newblock Nonenzymic template-directed synthesis on oligodeoxycytidylate
  sequences in hairpin oligonucleotides.
\newblock {\em J. Am. Chem. Soc.} {\bf 1992}, {\em 114},~317--322.

\bibitem{Wu:1992b}
Wu, T.; Orgel, L.
\newblock Nonenzymatic template-directed synthesis on hairpin oligonucleotides.
  3. Incorporation of adenosine and uridine residues.
\newblock {\em J. Am. Chem. Soc.} {\bf 1992}, {\em 114},~7963--7969.

\bibitem{Fer:2007}
Fernando, C.; Kiedrwoski, G.v.; Szathm{\'{a}}ry, E.
\newblock A stochastic model of nonenzymatic nucleic acid replication:
  {``}Elongators{''} sequester replicators.
\newblock {\em J. Mol. Evol.} {\bf 2007}, {\em 64},~572--585.

\bibitem{Mon:2010b}
Monnard, P.A.; D{\"{o}}rr, M.; L{\"{o}}ffler, P.
\newblock Possible role of ice in the synthesis of polymeric compounds.
\newblock  38th COSPAR Scientific Assembly, held 15-18 July 2010, Bremen,
  2010.

\bibitem{Kie:1986}
Kiedrowski, G.v.
\newblock A Self-replicating hexadeoxynucleotide.
\newblock {\em Angew. Chem.} {\bf 1986}, {\em 25},~932--935.

\bibitem{Sie:1994}
Sievers, D.; Kiedrowski, G.v.
\newblock Self-replication of complementary nucleotide-based oligomers.
\newblock {\em Nature} {\bf 1994}, {\em 369},~221--224.

\bibitem{Bag:1996}
Bag, B.G.; Kiedrowski, G.v.
\newblock Templates, autocatalysis and molecular replication.
\newblock {\em Pure \& App. Chem.} {\bf 1996}, {\em 68}.

\bibitem{Joy:1984}
Joyce, G.F.
\newblock Non-enzyme template-directed synthesis of RNA copolymers.
\newblock {\em Orig. Life Evol. Biosph.} {\bf 1984}, {\em 14},~613--620.

\bibitem{Lin:2009}
Lincoln, T.A.; Joyce, G.F.
\newblock Self-sustained replication of an RNA enzyme.
\newblock {\em Science} {\bf 2009}, {\em 323},~1229--1232.

\bibitem{Wil:1998}
Wills, P.; Kauffman, S.; Stadler, B.; Stadler, P.
\newblock Selection dynamics in autocatalytic systems: Templates replicating
  through binary ligation.
\newblock {\em Bulletin of Mathematical Biology} {\bf 1998}, {\em
  60},~1073--1098.

\bibitem{Roc:2006}
Rocheleau, T.; Rasmussen, S.; Nielson, P.E.; Jacobi, M.N.; Ziock, H.
\newblock Emergence of protocellular growth laws.
\newblock {\em Philos. Trans. R. Soc. B} {\bf 2007}, {\em 362},~1841--1845.

\bibitem{Sza:1989}
Sz{\'{a}}thmary, E.; Gladkih, I.
\newblock Sub-exponential growth and coexistence of non-enzymatically
  replicating templates.
\newblock {\em J. Theor. Biol.} {\bf 1989}, {\em 138},~55--58.

\bibitem{Kie:1991}
Kiedrowski, G.v.; Wlotzka, B.; Helbing, J.; Matzen, M.; Jordan, S.
\newblock Parabolic growth of a self-replicating hexadeoxynucleotide bearing a
  3{'}-5{'}-phosphoamidate linkage.
\newblock {\em Angew. Chem. Int. Ed.} {\bf 1991}, {\em 30},~423--426.

\bibitem{Lut:1998}
Luther, A.; Brandsch, R.; Kiedrowski, G.v.
\newblock Surface-promoted replication and exponential amplifcation of DNA
  analogues.
\newblock {\em Nature} {\bf 1998}, {\em 396},~245--248.

\bibitem{Zha:2006}
Zhang, D.Y.; Yurke, B.
\newblock A DNA superstructure-based replicator without product inhibition.
\newblock {\em Nat. Comput.} {\bf 2006}, {\em 5},~183--202.

\bibitem{Owc:1998}
Owczarzy, R.; Vallone, P.M.; Gallo, F.J.; Paner, T.M.; Lane, M.J.; Benight,
  A.S.
\newblock Predicting sequence-dependent melting stability of short duplex DNA
  oligomers.
\newblock {\em Biopolymers} {\bf 1998}, {\em 44},~217--239.

\bibitem{Blo:2000}
Bloomfield, V.A.; Crothers, D.M.; Tinoco, I.
\newblock {\em Nucleic Acids}; University Science Books: Sausalitos, CA, USA,
  2000.

\bibitem{Pol:1966b}
Poland, D.; Scheraga, H.A.
\newblock Occurrence of a Phase Transition in Nucleic Acid Models.
\newblock {\em J. Chem. Phys.} {\bf 1966}, {\em 45},~1456--1463.

\bibitem{Hut:2002}
Hutton, T.J.
\newblock Evolvable self-replicating molecules in an artificial chemistry.
\newblock {\em Artif. Life} {\bf 2002}, {\em 8},~341--356.

\bibitem{Smi:2003}
Smith, A.; Turney, P.; Ewaschuk, R.
\newblock Self-replicating machines in continuous space with virtual physics.
\newblock {\em Artif. Life} {\bf 2003}, {\em 9},~21--40.

\bibitem{Kle:1998}
Klenin, K.; Merlitz, H.; Langowski, J.
\newblock A Brownian Dynamics program for the simulation of linear and circular
  DNA and other wormlike chain polyelectrolytes.
\newblock {\em Biophysical Journal} {\bf 1998}, {\em 74},~780--788.

\bibitem{Tep:2005}
Tepper, H.L.; Voth, G.A.
\newblock A coarse-grained model for double-helix molecules in solution:
  Spontaneous helix formation and equilibrium properties.
\newblock {\em J. Chem. Phys.} {\bf 2005}, {\em 122}.

\bibitem{Dru:2000}
Drukker, K.; Schatz, G.C.
\newblock A model for simulating dynamics of DNA denaturation.
\newblock {\em J. Chem. Phys. B} {\bf 2000}, {\em 104},~6108--6111.

\bibitem{Fel:2007b}
Fellermann, H.; Rasmussen, S.; Ziock, H.J.; Sol{\'{e}}, R.
\newblock Life-cycle of a minimal protocell: a dissipative particle dynamics
  (DPD) study.
\newblock {\em Artif. Life} {\bf 2007}, {\em 13},~319--345.

\bibitem{Kub:1966}
Kubo, R.
\newblock The fluctuation-dissipation theorem.
\newblock {\em Rep. Prog. Phys.} {\bf 1966}, {\em 29}.

\bibitem{Ryc:1977}
Ryckaert, J.P.; Ciccotti, G.; Berendsen, H.J.C.
\newblock Numerical Integration of the Cartesian Equations of Motion of a
  System with Constraints: Molecular Dynamics of n-Alkanes.
\newblock {\em J. Comp. Phys.} {\bf 1977}, {\em 23},~327.

\bibitem{Ter:2002}
Teraoka, I.
\newblock {\em Polymer solutions {--} An introduction to physical properties};
  Wiley Interscience: New York, NY, USA,  2002.

\bibitem{San:1998}
SantaLucia, J.
\newblock A unified view of polymer, dumbbell, and oligonucleotide DNA
  nearest-neighbor thermodynamics.
\newblock {\em Proc. Nat. Acad. Sci. USA} {\bf 1998}, {\em 95},~1460--1465.

\end{thebibliography}

\end{document}